\documentclass[twocolumn,aps,showpacs,superscriptaddress]{revtex4}

	\usepackage{amssymb}
	\usepackage{amsfonts}
	\usepackage{amsmath}
	\usepackage{graphicx}
	\usepackage{bm}
	\usepackage{multirow}
 	\usepackage{graphicx}
	
	\newcommand{\vect}[1]{\boldsymbol{#1}}
	\newcommand{\op}[1]{\hat{\boldsymbol{#1}}}

	\newcommand{\hbn}{hBN} 
	
	\newcommand{\smfrac}[2]{\tfrac{ #1 }{#2}}

\begin{document}

\title{Zero-energy modes and valley asymmetry in the Hofstadter spectrum of bilayer graphene van der Waals heterostructures with hBN}

\author{Xi Chen}
\affiliation{National Graphene Institute, The University of Manchester, Manchester, M13 9PL, UK} 	
\author{J.~R.~Wallbank}
\affiliation{National Graphene Institute, The University of Manchester, Manchester, M13 9PL, UK}
\author{M.~Mucha-Kruczy\'{n}ski}
\affiliation{Department of Physics, University of Bath, Claverton Down, Bath, BA2 7AY, UK}
\author{E.~McCann}
\affiliation{Physics Department, Lancaster University, Lancaster, LA1 4YB, UK}
\author{V.~I.~Fal’ko}
\affiliation{National Graphene Institute, The University of Manchester, Manchester, M13 9PL, UK}

\begin{abstract}
We investigate the magnetic minibands of a heterostructure consisting of bilayer graphene (BLG)
and hexagonal boron nitride (hBN) by numerically diagonalizing a two-band Hamiltonian that describes
electrons in BLG in the presence of a moir\'e potential.
Due to inversion-symmetry breaking characteristic for the moir\'e potential,
the valley symmetry of the spectrum is broken, but despite this, the zero-energy Landau level in BLG survives, albeit with reduced degeneracy.
In addition, we derive effective models for the low-energy features in the magnetic minibands
and demonstrate the appearance of secondary Dirac points in the valence band, which we confirm by numerical simulations.
Then, we analyze how single-particle gaps in the fractal energy spectrum produce a sequence of incompressible states
observable under a variation of carrier density and magnetic field.
\end{abstract}
\maketitle

\section{Introduction}
Long-period moir\'e patterns are characteristic of well-aligned graphene heterostructures with hBN. They offer an experimentally-viable route \cite{ponomarenko_nature_2013, dean_nature_2013, hunt_science_2013} to observe fractal magnetic miniband spectra generic for electrons in two-dimensional superlattices in strong magnetic fields \cite{brown_physrev_1964, zak_physrev_1964a, zak_physrev_1964b}.
These fractal miniband spectra, also known as Hofstadter's butterflies \cite{hofstadter_prb_1976}, reflect the fact that the translational symmetry of a lattice, suppressed by the presence of a magnetic vector potential in the electron's Schr\"odinger equation, is restricted to magnetic field values $B_{p/q}=(p/q)\phi_{0}/S$ corresponding to a rational fraction $p/q$ of magnetic flux quantum $\phi_0=h/e$ per unit cell area $S$ of the lattice \cite{brown_physrev_1964, zak_physrev_1964a, zak_physrev_1964b}.
The generic features of fractal magnetic miniband spectra have been observed \cite{ponomarenko_nature_2013, hunt_science_2013, Yu_NatPhys_2014, wang_science_2015} and modeled \cite{wallbank_prb_2013, chen_prb_2014, Yu_NatPhys_2014,diez_prl_2014,moon_prb_2014,chizhova_prb_2014,wallbank_adp_2015,slotman_prl_2015} in monolayer graphene heterostructures with {\hbn}, where the properties of Dirac electrons at the conduction/valence band edge allow for a straightforward interpretation of experimental observations, and have also been studied in slightly misaligned pairs of graphene flakes \cite{bistritzer_prb_2011, Santos_PRL_2007}.
Modeling of the fractal magnetic miniband spectra of a bilayer graphene/hBN heterostructure is much more limited \cite{moon_prb_2014}, even though one of the first observations of this phenomena was in this system~\cite{dean_nature_2013}.

In this Article, we study fractal magnetic minibands in bilayer graphene (BLG) subject to a moir\'e superlattice perturbation due to an hBN underlay.
The low-energy Hofstadter butterfly spectra of a BLG-{\hbn} heterostructure is dominated by bands related to the degenerate ``zero-energy'' Landau level (LL) states $n=0$ and $n=1$ of unperturbed BLG \cite{mccann_prl_2006}.
We use a symmetry-based approach to model the influence of {\hbn} on electrons in BLG, and we show that the spectra in BLG-{\hbn} is a superposition of two very different miniband spectra associated with electrons in opposite valleys (Brillouin zone corners) of BLG's band structure:
the miniband spectrum in one valley is only weakly broadened by the superlattice potential and incorporates one unperturbed $n=0$ LL, while the miniband spectrum in the other valley is widely broadened.
This valley asymmetry arises from spatial inversion symmetry breaking produced by the fact that the moir\'e perturbation only directly affects one of the two BLG layers.
Since the zero energy LL states reside on different layers in opposite valleys, in one valley they are strongly influenced by the moir\'e perturbation, but in the other valley they are not.

These qualitative features agree with the results of the tight-binding model of Moon and Koshino \cite{moon_prb_2014}. In addition, we characterize different types of moir\'e superlattice perturbations including the possibilities that the perturbation creates potential asymmetry between the carbon atoms in BLG or introduces a spatial modulation of the nearest-neighbor carbon-carbon hopping amplitude. Hence, we find that the Dirac point at the conduction-valence band edge can be either gapless or gapped, and that a secondary Dirac point can appear in the valence band of BLG-{\hbn}.
Our model Hamiltonian is presented in Section~\ref{S:ham} and its numerical diagonalization in the presence of a magnetic field, including methodology, results and discussion, is described in Section~\ref{S:numerics}, where we also derive simple effective Hamiltonians to describe low-energy features in the magnetic minibands
and, also, an effective Hamiltonian to describe the secondary Dirac point. In Section~\ref{S:map}, we show how gaps in the fractal energy spectrum are manifest as observable incompressible states under a variation of carrier density and magnetic field.

\section{Moir\'e superlattice Hamiltonian}\label{S:ham}

\subsection{Four band model}
To describe the sublattice ($A/B$) and bottom/top ($1/2$) layer composition of electron states in Bernal-stacked BLG on a {\hbn} underlay, we write down their Hamiltonian \cite{mccann_prl_2006} as a $4\times 4$ matrix,
\begin{align}
\label{eqn:4x4Hamiltonian}
&\op{H}_\xi \!\!=\! \!\left(\!\!
\begin{array}{cc}
v \op{p} \!\cdot \!\vect\sigma + \op{M}_{\xi} & \gamma_1(\xi \sigma_1\!-\! i\sigma_2)/2\\
(\gamma_1(\xi \sigma_1\!-\! i\sigma_2)/2)^{\dagger } & v \op{p} \!\cdot \!\vect\sigma
\end{array}
\!\!\right)
\\
& \op{M}_\xi= v b  u_0  f_+ \!
 + \xi v b \sigma_3 u_3   f_-
 +\xi v \vect \sigma \!\cdot\! \left[\vect l_z\!\times\!\nabla\! \left( u_1  f_-\right) \right]\!.
 \nonumber
\end{align}
Here, we use the basis of Bloch functions $(\phi_{A_1},\phi_{B_1},\phi_{A_2},\phi_{B_2})$ for valley $K$ ($\xi=1$), $(\phi_{B_1},-\phi_{A_1},\phi_{B_2},-\phi_{A_2})$ for valley $K'$ ($\xi=-1$), and the $2\times 2$ Pauli matrices $\sigma_{1,2,3}$ act in the space of sublattice components.
The term $v \op{p} \!\cdot \!\vect\sigma$ on the diagonal takes into account the electrons' Dirac spectrum in each layer where $v\simeq10^8\,$cm$\,$s$^{-1}$ \cite{deacon_prb_2007} is the Dirac velocity, $ \hat{\vect p} = -i \nabla + e \vect A$, $[\nabla\times\vect A]_z=-B$, $\vect \sigma=(\sigma_x,\sigma_y)$, and $\hbar=1$, while
$\gamma_1\simeq 0.4\,$eV \cite{kuzmenko_prb_2009} is the inter-layer hopping between $A_2$ and $B_1$ sublattices.

The term $\op{M}_\xi$ in Eq.~(\ref{eqn:4x4Hamiltonian}) accounts for the moir\'e superlattice perturbation produced by the {\hbn} substrate \cite{wallbank_prb_2013, chen_prb_2014}, and is applied only to the layer of the bilayer which is closest to the {\hbn} (layer `$1$') \cite{mucha-kruczynski_prb_2013}.
Within the term $\op{M}_\xi$, the functions $f_\pm =\!\sum_{m}\!(\pm1)^{m+\frac{1}{2}}e^{i\vect b_m\cdot\vect r}$, with $m=0,1,...,5$, are written using the shortest non-zero Bragg vectors of the moir\'e superlattice $\vect{b}_m=\op{R}_{m\pi/3} [1-(1+\delta)^{-1} \op{R}_{\theta}] (0,\smfrac{4\pi}{\sqrt{3}l})$, where $\op{R}_\varphi$ describes anti-clockwise rotation by angle $\varphi$, $\delta\approx1.8\%$ takes into account the relative lattice mismatch between graphene and {\hbn} \cite{xue_natmater_2011}, $\theta$ is the misalignment angle between the two hexagonal lattices and $l=2.46\,$\AA$\,$ is the lattice constant of graphene.
For $\theta\ll1$, $b=|\vect b_m|\approx \smfrac{4\pi}{\sqrt{3}l}\sqrt{\delta^2+\theta^2}$ so that the energy scale $vb$ obtains its minimum value of $vb \approx 0.35\,$eV at $\theta=0$.
The strength of the various terms included in $\op{M}_\xi$ are controlled by separate dimensionless parameters, $u_{i=0,3,1}$, which describe in turn an electrostatic potential which
does not distinguish between the two carbon sublattices ($u_0$), a sublattice-asymmetric part of the potential ($u_3$), and a spatial modulation of the nearest-neighbor carbon-carbon hopping amplitude ($u_1$).

\subsection{Two band model}

\begin{figure*}[htbp]
\centering
\includegraphics[width=.9 \textwidth]{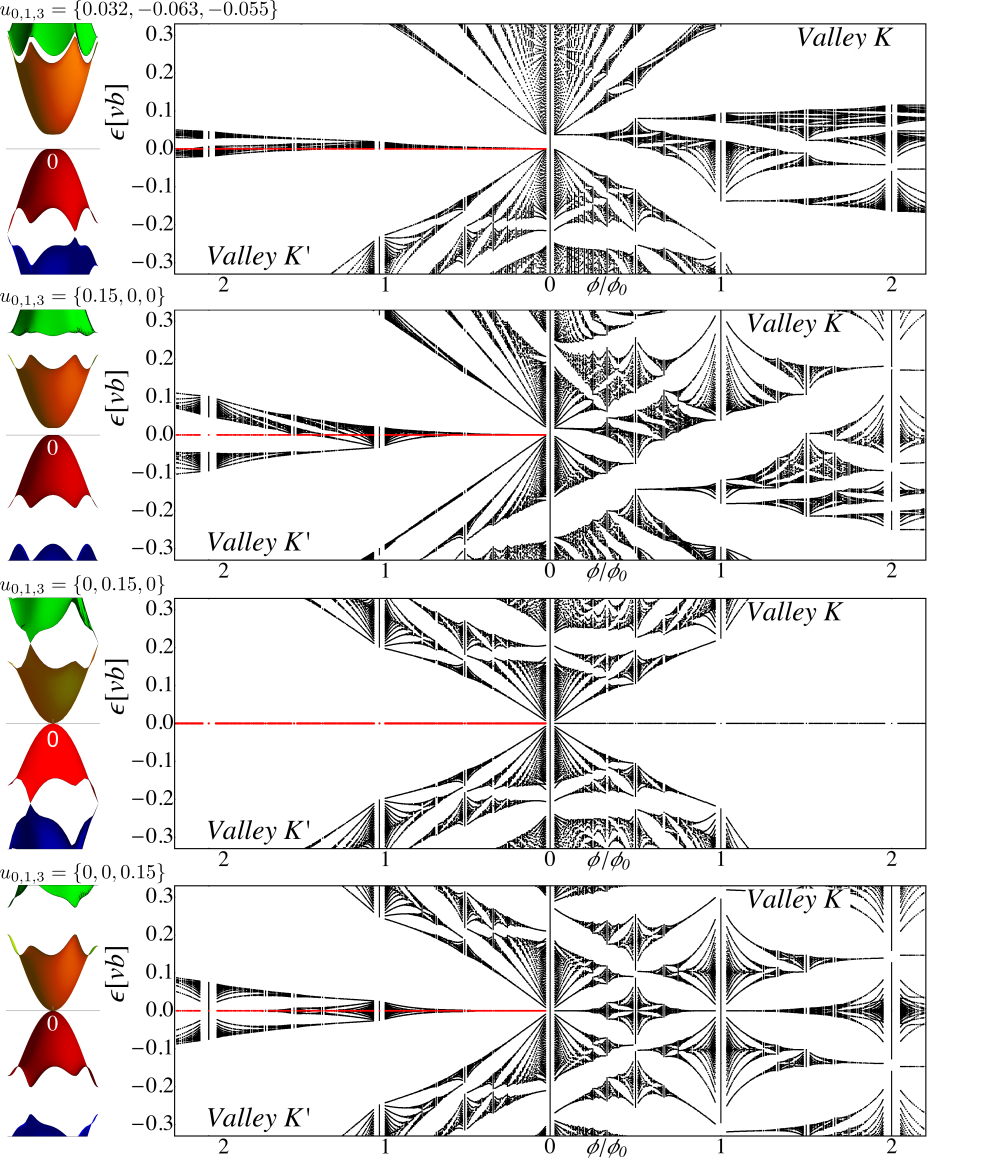}
\caption{
The $B=0$ minibands for the labeled superlattice perturbation parameters and $\theta=0$ (left panels), and the bandwidths of the magnetic minibands, shown separately for the $K'$ and $K$ valleys, for the same superlattice parameters (central panels)
A red line is used to indicate the unperturbed zero energy Landau level in the $K'$-valley.
}
\label{fig:magnetic_minibands}
\end {figure*}

One of the most interesting features of electrons in BLG at strong magnetic fields is the degeneracy of two orbital LLs, with $n=0$ and $n=1$, which appear at $\epsilon=0$, the edge between the valence and conduction band. The mixing of these degenerate LLs by the moir\'e superlattice potential determines the main features of the lower-energy part of the magnetic miniband spectrum, shown in Fig.~\ref{fig:magnetic_minibands} for several different  choices of moir\'e perturbation parameters $u_i$.
These low-energy electron states in BLG can be described \cite{mccann_prl_2006} using a simplified two-band Hamiltonian, which can be obtained from  Eq.~\eqref{eqn:4x4Hamiltonian} by a Schrieffer-Wolf projection onto the basis of Bloch states residing on $A_1$ and $B_2$ sublattices. For a BLG-{\hbn} heterostructure, such a projection results in \cite{mucha-kruczynski_prb_2013},
\begin{align}\label{eqn:2x2Hamiltonian}
& \boldsymbol{\tilde{H}}_{\xi}=
-\xi\frac{v^2}{\gamma_1} \left(\begin{array}{cc} 0 &\left(\op{\pi}^{\dagger}\right)^{2}\\ \op{\pi}^2 & 0\end{array}\right)
 + \boldsymbol{\tilde{M}}_{\xi}, \\
& \boldsymbol{\tilde{M}}_{1}=\left(
\begin{array}{cc}
 v b g_+(r) & \smfrac{b v^2}{\gamma_1}h_+^*(r)\op{\pi}^{\dagger } \\[0.3em]
\smfrac{b v^2}{\gamma_1}\op{\pi} h_+(r) & \smfrac{v^3 b}{\gamma_1^2}\op{\pi} g_-(r)\op{\pi}^{\dagger}
\end{array}
\right),\nonumber\\
& \boldsymbol{\tilde{M}}_{\text{-}1}=\left(
\begin{array}{cc}
 \smfrac{v^3 b}{\gamma_1^2} \op{\pi}^{\dagger }g_-(r)\op{\pi} & \smfrac{b v^2}{\gamma_1}\op{\pi}^{\dagger }h_-(r) \\[0.3em]
 \smfrac{b v^2}{\gamma_1}h_-^*(r)\op{\pi} & v b g_+(r)
\end{array}
\right),\nonumber \\
& g_{\pm }(r)=\sum _{m} e^{i b_m.r} (u_0\pm i u_3(-1)^m),\nonumber\\
& h_{\pm}(r)= \pm i u_1  \sum _m(-1)^me^{i b_m.r}\left(b_m^x \pm i b_m^y \right)/b. \nonumber
\end{align}
Hamiltonian $\boldsymbol{\tilde{H}}_{\xi}$ is written in the basis of the Bloch states $(\phi_{A_1},\phi_{B_2})$ for the $K$ valley and $(\phi_{B_2}, -\phi_{A_1})$ for the $K'$ valley using $\op{\pi}=\op{p}_x+i\op{p}_y$, and $\vect b_m=(b^x_m,b^y_m)$

Examples of the zero magnetic field $K$-valley bandstructure of moir\'e perturbed BLG with $\theta=0$ are displayed in the left panels of Fig.~\ref{fig:magnetic_minibands}.
These spectra were calculated by numerical diagonalization of the Heisenberg matrix constructed from Hamiltonian \eqref{eqn:2x2Hamiltonian} in a basis of unperturbed plane wave states.
The corresponding dispersion in the $K'$-valley is obtained using the relation, $\epsilon_{K'}(\vect k)=\epsilon_{K}(-\vect k)$, which follows from the time-reversal symmetry of Hamiltonians \eqref{eqn:4x4Hamiltonian} and \eqref{eqn:2x2Hamiltonian} for $B=0$.

Each panel in Fig.~\ref{fig:magnetic_minibands} corresponds to a different set of moir\'e perturbation parameters $\{ u_0 , u_1 , u_3 \}$.
The choice of parameters used in the top panel corresponds to the predictions of a pair of microscopic models, one based on hopping between graphene
carbon atoms and hBN atoms, the other on scattering of graphene electrons by the quadropole
electric moments of nitrogen atoms. Interestingly, these two models predict the same relative values for the moir\'e potential parameters \cite{wallbank_prb_2013},
\begin{align}
 u_{i=0,1,3}=V\left\{1/2,-\delta/\sqrt{\delta^2+\theta^2},-\sqrt{3}/2\right\} ,
\label{eq:model}
\end{align}
where $V$ depends upon the specific microscopic parameters used to describe the hBN underlay. Here we take $V=0.063$, chosen to make the moir\'e perturbation strong enough that an almost gapless secondary Dirac point (sDP) is produced between the red and blue minibands in the valence band of the upper left panel of Fig.~\ref{fig:magnetic_minibands}. Signatures of this feature were observed experimentally in Ref.~\cite{dean_nature_2013}.

The other panels in Fig.~\ref{fig:magnetic_minibands} illustrate the influence of each $u_i$ parameter taken individually and they exemplify three additional scenarios: for $u_{0,1,3}=\{0.15 , 0, 0 \}$, the original Dirac point is gapped and there is no sDP (in the two valence bands closest to zero energy); for $u_{0,1,3}=\{ 0, 0.15 , 0 \}$, the original Dirac point and the sDP are both gapless (and there is a sDP in the conduction band); for $u_{0,1,3}=\{0, 0, 0.15\}$, the original Dirac point is gapless and there is no sDP.

\section{
The BLG-hBN heterostructure in a magnetic field studied using the two band model}\label{S:numerics}

\subsection{Methodology}
Here we describe the numerical diagonalization of Hamiltonian \eqref{eqn:2x2Hamiltonian} in the presence of a magnetic field.
To simplify this calculation, we take account of the hexagonal symmetry of the moir\'e pattern and use a non-orthogonal coordinate system \cite{chen_prb_2014} $\vect{r}=(x_1\vect{a_{1}}+x_2\vect{a_{2}})/a$,
where
$\vect a_1=4\pi \vect b_1\times \vect{\hat{l}}_z/(\sqrt3 b^2)$ and $\vect a_2=4\pi \vect{\hat{l}}_z \times \vect b_5/(\sqrt3 b^2)$ are direct moir\'e lattice vectors and $a=|\vect a_1|$.
In this basis the Landau gauge has the form $\vect{A}= B x_1(\vect{a_{1}}-2\vect{a_{2}})/(\sqrt{3}a)$ which leads to
\begin{align*}
\op{\pi} = \smfrac{-2}{\sqrt{3}}\left[ \partial_{x_1} \tau +  (\partial_{x_2}-  i\sqrt{3} e B x_1/2)\tau^* \right],
\end{align*}
where $\tau=e^{i2\pi/3}$, and $e$ is the electron charge. For the wave-vector space, this determines $\vect{k}=k_{1}\vect{k}_1 +k_{2}\vect{k}_2 $ with $\vect{k}_1 =2\vect b_5/(\sqrt 3 b) $ and  $\vect{k}_2 =2\vect b_1/(\sqrt 3 b) $.
Hence, we employ the basis set of magnetic oscillator functions,
\begin{align}\label{eq:mag_oscil_funcs}
& \varphi_{n}(k_{2}) = \frac{\sqrt{ 3  }e^{ik_{2}x_{2}}  e^{-\frac{z^2}{2}- \frac{i z^2}{2 \sqrt{3}}}}{\sqrt{n!2^{(n+1)} \lambda_B\sqrt{\pi}}} \mathbb{H}_n(\!z\!), \\
& z=\frac{\sqrt{3} x_1}{2\lambda_B}- k_2\lambda_B, \quad
\lambda_B\!=\!1/\sqrt{|eB|}, \nonumber \\
& \op{\pi}\varphi_n(k_2)=-\tau \lambda_B^{-1}\sqrt{2n}\varphi_{n-1}(k_2),\nonumber
\end{align}
where $\mathbb{H}_n(z)$ are the Hermite polynomials.
For free electrons in BLG, the LL spectrum contains two degenerate states at $\epsilon=0$,  and pairs of conduction/valence band states at
$\epsilon_{n\ge 2}^{\alpha}=\alpha\sqrt{n(n-1)}/m\lambda_{B}^{2}$, $\alpha = \pm 1$,
\begin{align} \label{eqn:basis function}
&\psi^0_{n=0,1}(k_{2})= \frac{1}{\sqrt{L}} \left(\begin{array}{c}\varphi_{n}(k_{2}) \\0\end{array}\right) \, ,\\
&\psi_{|n|\ge2}^{\alpha}(k_{2}) = \frac{1}{\sqrt{2L}} \left(\begin{array}{c}\varphi_{n}(k_{2}) \\\text{-}\alpha \tau \varphi_{n\text{-}2}(k_{2})\end{array}\right) \, . \label{eqn:basisfnall}
\end{align}

In general, the magnetic vector potential $\vect{A}(x_{1})$ breaks the symmetry of the Hamiltonian with respect to translations of the moir\'e superlattice.
However, for magnetic flux $\phi=S B=\frac{p}{q}\phi_0$ where $p$ and $q$ are co-prime natural numbers
and $S$ is the unit cell area of the moir\'e superlattice, translational symmetry is restored.
Because of this, we consider a unit cell of the magnetic superlattice that is $q$ times larger than the unit cell of the moir\'e superlattice in both directions (hence its area is $q^2$ times larger) \cite{brown_physrev_1964,chen_prb_2014}.
The magnetic translational group $G=\{\Theta_{\vect X}\equiv e^{ieBm_1qa\sqrt3x_2/2} T_{\vect X}, \vect X=m_1q\vect a_1+m_2q\vect a_2\}$, where $T_{\vect X}$ describes geometrical translations and $m_1$ and $m_2$ are integers, commutes with Hamiltonian \eqref{eqn:2x2Hamiltonian} and is isomorphic to the group of geometrical translation, so that its eigenstates form a plane wave basis $\Theta_{\vect X} |n^{\alpha}_{jt}(\vect{k})\rangle = e^{i\vect{k}\cdot\vect{X}}|n^{\alpha}_{jt}(\vect{k})\rangle$.
Bloch functions $|n^{\alpha}_{jt}(\vect{k})\rangle$
exist in the magnetic Brillouin zone which is $q^2$ times smaller than that of the moir\'e superlattice \cite{brown_physrev_1964}
and in which magnetic minibands are $q$-fold degenerate.
For states with momentum $\vect{k}$ within this magnetic Brillouin zone,
\begin{align}\label{eq:bloch_wf}
|n^{\alpha}_{jt}(\vect{k})\rangle\!\! =\!\! \frac{1}{\sqrt N}\!\!\sum_{s=\text{-}N/2}^{N/2} \!\!\! e^{\text{-}ik_1 qas} \psi_n^{\alpha}(k_2\!+\!\smfrac{\sqrt3}{2}b[ps\!+\!j\!+\!t\smfrac{p}{q}]),
\end{align}
where $N \rightarrow \infty$, and $j=0,\cdots,p-1$ indexes the magnetic sub-bands, and $t=0,\cdots,q-1$ indexes the above mentioned $q$-fold degeneracy
Without loss of generality, we set $t=0$ and omit it, using the notation $|n^{\alpha}_{j}(\vect{k})\rangle$, from now on.
In order to obtain the energy dispersion, we calculate the matrix elements of the perturbation,  $\langle n^{\alpha}_j(\vect{k})|\boldsymbol{\tilde{H}}^B_{\xi}|\tilde{n}^{\tilde{\alpha}}_{\tilde{j}}(\tilde{\vect{k}})\rangle$, in this basis and diagonalize the resulting matrix \cite{chen_prb_2014}.

\subsection{Results and discussion}
The main panels of Fig.~\ref{fig:magnetic_minibands} show the magnetic spectrum of the BLG-{\hbn} heterostructure for the four choices of moir\'e perturbation described above.
For small flux, the magnetic miniband spectra can be traced to the sequence of Landau levels for moir\'e perturbed BLG.
Near zero energy, the gap at the original Dirac point is seen in the magnetic spectra of the top two panels.
The presence or absence of a secondary Dirac point (depending on the particular parameter set) is also clearly reflected in the magnetic spectra at small flux at the corresponding energy.
At a higher flux, the Landau levels broaden and split, forming a fractal pattern, with the most striking features around zero energy.

For all parameter sets, the valley symmetry of the spectrum, preserved in the absence of a magnetic field, is lifted.
This is because the moir\'e perturbation only affects the layer of BLG that is adjacent to
the {\hbn} and, thus, it breaks inversion symmetry. In conjunction with time-inversion symmetry breaking by a magnetic field, this allows the energy spectra for electrons in valleys $K$ and $K'$ to be different \cite{moon_prb_2014}.
In particular, note that, in the absence of the perturbation, the distribution of the wave function among the layers is, for a given Landau level, exactly inverted in the two valleys, Eqs.~(\ref{eqn:basis function},\ref{eqn:basisfnall}).
The moir\'e potential affects only the wave function component in the
layer adjacent to {\hbn}, hence breaking the layer symmetry and leading to valley-asymmetric spectra with a gap.

Importantly, the spectrum in the valley for which the wave function sits on the layer further from the substrate contains a zero-energy Landau level completely decoupled from the rest of the spectrum (shown as a red line in the $K'$-valley in Fig.~\ref{fig:magnetic_minibands}).
The spinor structure Eq.~\eqref{eqn:basis function} of the two zero-energy states, $n=0,1$, for unperturbed electrons shows that the states in valley $K$ ($K'$) are localized on the bottom (top)
graphene layer only \cite{mccann_prl_2006}.
Since the moir\'e perturbation does not directly influence the top layer,
it does not couple the $n=0$ state in the $K'$ valley to any other state and, thus, the $n=0$ state in this valley always persists.
This coexistence of a zero-energy LL and a fractal spectrum of magnetic sub-bands creates a unique opportunity to observe the interplay between electron-electron interaction and Hofstadter's quantization \cite{ghazaryan_prb_2015, wang_science_2015}.

Each of the panels in Fig.~\ref{fig:magnetic_minibands} display further features of interest.
For the parameter set $u_{0,1,3}=\{0.032, -0.063, -0.055\}$ (upper panel), the $B=0$ miniband spectrum displays a slightly gapped sDP found in the first valence miniband, which produces a sequence of LL including tilted zeroth LLs, at the corresponding energy and weak magnetic fields ($\epsilon\approx-0.22\,vb$ $\phi/\phi_0\lesssim0.2$). We shall discuss the sDP further in Section~\ref{S:sdp} (a similar feature is present in the lower middle panel).
For potential modulation taken in isolation, $u_{0,1,3}=\{0.15, 0, 0\}$, the band structure obeys six-fold rotation symmetry in each valley (left upper middle panel), in contrast to all other band structure images which only display symmetry under three-fold rotation.
For the spatial modulation of the carbon-carbon hopping amplitude, $u_{0,1,3}=\{0, 0.15, 0\}$ (lower middle panels), the spectrum at both zero magnetic field and finite magnetic field obeys electron-hole symmetry, and the zeroth LL and first LL of the original Dirac point are completely degenerate in both valleys (discussed in Section \ref{S:low-e}).
For the sublattice-asymmetric potential, $u_{0,1,3}=\{0, 0, 0.15\}$ (lower panel), the bands at zero magnetic field are symmetric under the operation which combines electron-hole-reflection and a rotation by $\pi/3$, which is reflected in the electron-hole symmetry of the magnetic miniband widths in a magnetic field.
Additionally for this perturbation,  we find that the magnetic miniband structure around $\epsilon=0$ can form gapless linear or quadratic Dirac points, which produce sequences of LLs in the magnetic miniband spectrum, best visible around $\phi/\phi_0=2$ in the $K'$-valley and  $\phi/\phi_0=1$ in the $K$-valley (also see Section \ref{S:low-e}).

\subsection{Effective model of low energy features in the magnetic minibands}\label{S:low-e}
\begin{figure}[htbp]
\centering
\includegraphics[width=.49 \textwidth]{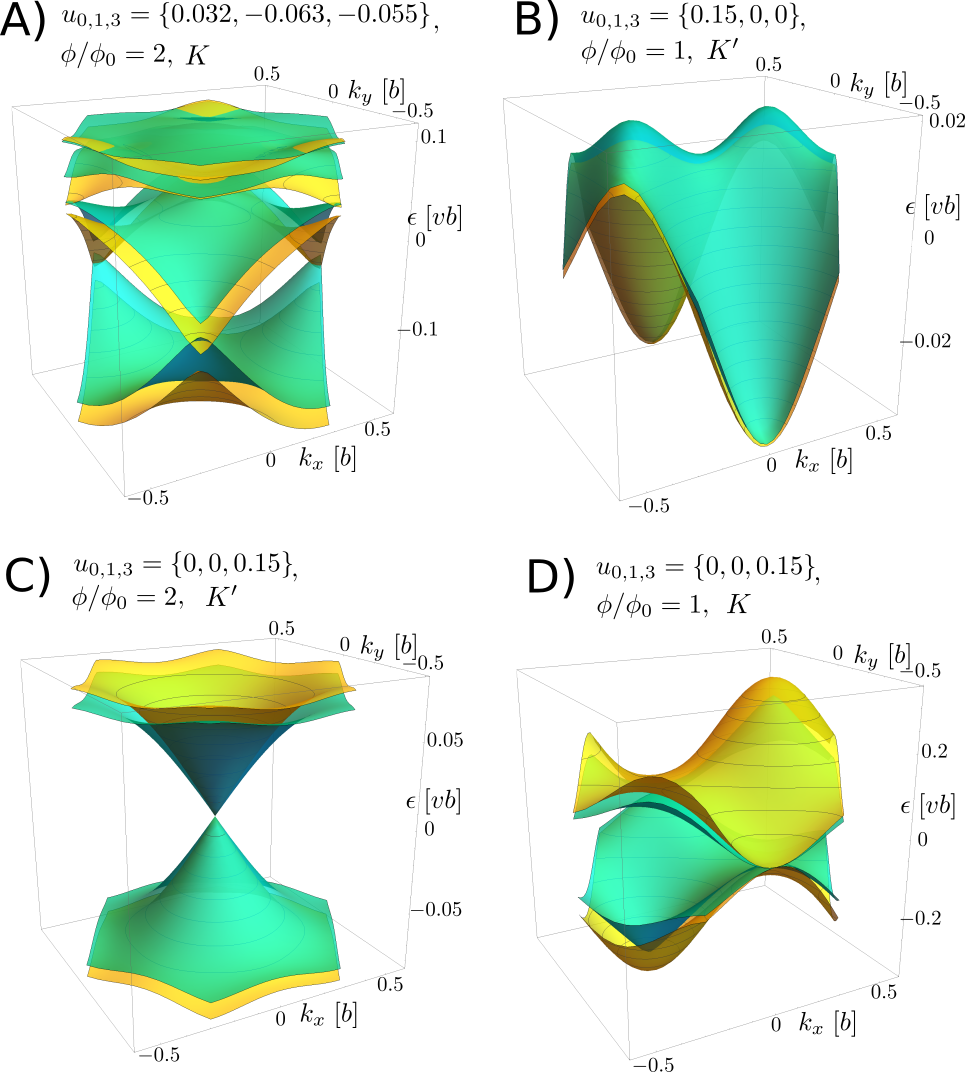}
\caption{
The full $(k_x,k_y)$-dispersion of magnetic minibands obtained by either diagonalizing effective Hamiltonians \eqref{eq:Heff1}-\eqref{eq:Heff2} (yellow), or the full numerical diagonalization of Hamiltonian \eqref{eqn:2x2Hamiltonian} (turquoise).}
\label{fig:low_energy_magnetic_minibands}
\end {figure}
The main features of the broadened magnetic miniband spectrum around zero energy at a simple fraction $\phi/\phi_0=1/q$ can be described by truncating the basis of Bloch functions Eq.~\eqref{eq:bloch_wf}
to the zeroth and first LL only, yielding effective Hamiltonians,
\begin{align}\label{eq:Heff1}
 &\boldsymbol{  {\mathbb{H}}}_{K'}
= 0\oplus vb D   \omega  (u_0 f_0 - u_3  f_1) \, ,\\
&\boldsymbol{  {\mathbb{H}}}_{K}
=vb D   u_0 \left(
\begin{array}{cc}
 2 f_0 & \sqrt{2 b l_0 } f_2^*\\
 \sqrt{2 b l_0 } f_2& (2-b l_0 )f_0
\end{array}
\right)  \nonumber\\
&     \qquad + vb D   u_3  \left(
\begin{array}{cc}
 2 f_1 & \sqrt{2 b l_0 } f_3^*\\
 \sqrt{2 b l_0 } f_3  & (2-b l_0 )f_1
\end{array}
\right) \, ,\nonumber\\
 &f_0=c_1 +c_2 +(\text{-}1)^{q}  c_{12}, \
f_1=\text{-}s_1 +s_2 +(\text{-}1)^{q}  s_{12} \, ,\nonumber\\
&f_2=\tau^* \!s_1 \!-\!s_2 \!-\tau (\text{-}1)^{q}  s_{12} ,
f_3=\tau^* \!c_1 \!+\!c_2 \!+\tau (\text{-}1)^{q}  c_{12}\, .\nonumber
\end{align}
Here $\boldsymbol{  {\mathbb{H}}}_{K/K'}$ is used for the $K/K'$-valley and written using the basis ($|0^0_{0,0}\rangle$,$|1^0_{0,0}\rangle$).
Also $l_0  = (4 \pi /\sqrt{3} b) (\phi _0/\phi )$,
$\omega  =(\sqrt{3}b^2 v^2 /4\pi \gamma_1^2) (\phi/\phi_0)$,
$c_1=\cos(l_0   k_1)$, $c_2=\cos(l_0   k_2)$, $c_{12}=\cos(l_0   (k_1-k_2))$, $s_1=\sin(l_0   k_1)$, $s_2=\sin(l_0   k_2)$ and $s_{12}=\sin(l_0   (k_1-k_2))$.
The factor $D=\exp [-(\pi/\sqrt{3})(\phi_0/\phi)]$ leads to rapid broadening for $\phi<\phi_{0}$ which slows down for $\phi>\phi_{0}$.
In $\boldsymbol{  {\mathbb{H}}}_{K'}$ the symbol $\oplus$ denotes a direct sum, where the state $|0^0_{0,0}\rangle$ is decoupled from the rest of the spectrum.

Similarly, for simple fractions $\phi/\phi_0=1/({\cal{N}}+1/2)$, with integer ${\cal{N}}$,
the spectrum around zero energy may be described by effective Hamiltonians
\begin{align}\label{eq:Heff2}
&\boldsymbol{ \tilde{\mathbb{H}}}_{K' }
= 0_{2,2}\oplus vb D   \omega \left( u_0 M_1 -   u_3 M_2\right) \, , \\
&\boldsymbol{ \tilde{\mathbb{H}}}_{K}
=vb D   u_0 \left(\begin{array}{cc}2 M_1 & \sqrt{2 b l_0 }  M_3^{\dagger} \\ \sqrt{2 b l_0 }   M_3 & (2-bl_0 )M_1\end{array}\right) \nonumber\\
&\qquad + vb D   u_3 \left(\begin{array}{cc} 2 M_2 & \sqrt{2 b l_0 }   M_4^{\dagger} \\ \sqrt{2 b l_0 }  M_4 & (2-b l_0 ) M_2\end{array}\right)\, , \nonumber
\end{align}
where the $2 \times 2$ block matrices are
\begin{align}
&M_1= \left(
\begin{array}{cc}
  c_2 &  e^{\text{-}i l_0   k_1} [c_1 -i c_{12}] \\
  e^{i l_0   k_1} [c_1 +i c_{12}] & - c_2
\end{array}
\right) \nonumber \, ,\\
&M_2=\left(
\begin{array}{cc}
  s_2 &  -e^{\text{-}i l_0   k_1} [s_1 +i s_{12}] \\
  -e^{i l_0   k_1} [s_1 -i s_{12}] &  -s_2
\end{array}
\right) \nonumber \, ,\\
&M_3= \left(\begin{array}{cc}
-s_2 & e^{\text{-}i l_0   k_1}[\tau^*s_1 -i \tau s_{12}] \\
 e^{i l_0   k_1}  [\tau^*s_1 +i  \tau s_{12}] & s_2
 \end{array}\right)\nonumber \, ,\\
&M_4=\left(\begin{array}{cc}
 c_2 & e^{\text{-}i l_0   k_1} [\tau^*c_1 +i \tau c_{12}] \\
 e^{i l_0   k_1}  [\tau^*c_1 -i \tau c_{12}] & -c_2 \end{array}\right) \, .\nonumber
 \end{align}
Here we use the basis $|0^0_{0,0}\rangle$,$|0^0_{1,0}\rangle$,$|1^0_{0,0}\rangle$,$|1^0_{1,0}\rangle$
and $0_{2,2}$ is the $2\times2$ zero matrix.

Figure \ref{fig:low_energy_magnetic_minibands} shows the excellent agreement between the result of diagonalizing effective Hamiltonians \eqref{eq:Heff1}-\eqref{eq:Heff2} (yellow), with the fully numerical diagonalization of Hamiltonian \eqref{eqn:2x2Hamiltonian} (turquoise), for the various choices of magnetic field, valley, and superlattice parameters $u_i$ indicated on the figure.
The computed dispersion surfaces display a wide array of possible forms, including the possibility of gapless linear or quadratic Dirac points for the superlattice perturbation $u_{i=0,1,3}=\{0,0,0.15\}vb$. These features are found either in the $K'$-valley whenever $p$ in $\phi / \phi_{0} = p/q$ is even (Fig.~\ref{fig:low_energy_magnetic_minibands}C), or in the $K$-valley whenever $p$ in $\phi / \phi_{0} = p/q$ is odd (Fig.~\ref{fig:low_energy_magnetic_minibands}D).

For the perturbation with $u_{i=0,1,3}=\{0,0.15,0\}vb$ two zero-energy LLs of Hamiltonian \eqref{eqn:2x2Hamiltonian} remain unperturbed (lower middle panel in Fig.~\ref{fig:magnetic_minibands}), which is reflected in the fact that all matrix elements in Hamiltonians \eqref{eq:Heff1}-\eqref{eq:Heff2} vanish.
In this case, Hamiltonian \eqref{eqn:2x2Hamiltonian} obeys the ``electron-hole'' symmetry $\sigma_z \vect{\tilde{H}}_\xi \sigma_z=-\vect{\tilde{H}}_\xi$, which implies that its matrix elements must obey $\langle n^{\alpha}_j(\vect{k})|\vect{\tilde{H}}_\xi| \tilde{n}^{\alpha}_{ \tilde{j}}(\vect{k})\rangle = -\langle n^{\text{-}\alpha}_j(\vect{k})|\vect{\tilde{H}}_\xi| \tilde{n}^{\text{-}\alpha}_{\tilde{j}}(\vect{k})\rangle$.
Consequently, the resulting Heisenberg matrix has at least $2p+1$ linearly dependent rows, resulting in zero-energy eigenvalues which are $2p$-fold degenerate.
As the $u_1$ only perturbation can be considered to be a periodic pseudo-magnetic field \cite{wallbank_prb_2013}, it could alternatively be created using an Abrikosov lattice, i.e., a vortex structure generated from a type-II superconductor in a magnetic field. Such a system is also expected to possess degenerate zero energy eigenvalues \cite{Chen_PRB_2016}.

\subsection{Effective models for the secondary Dirac points}\label{S:sdp}
\begin{figure}[htbp]
\centering
\includegraphics[width=.4 \textwidth]{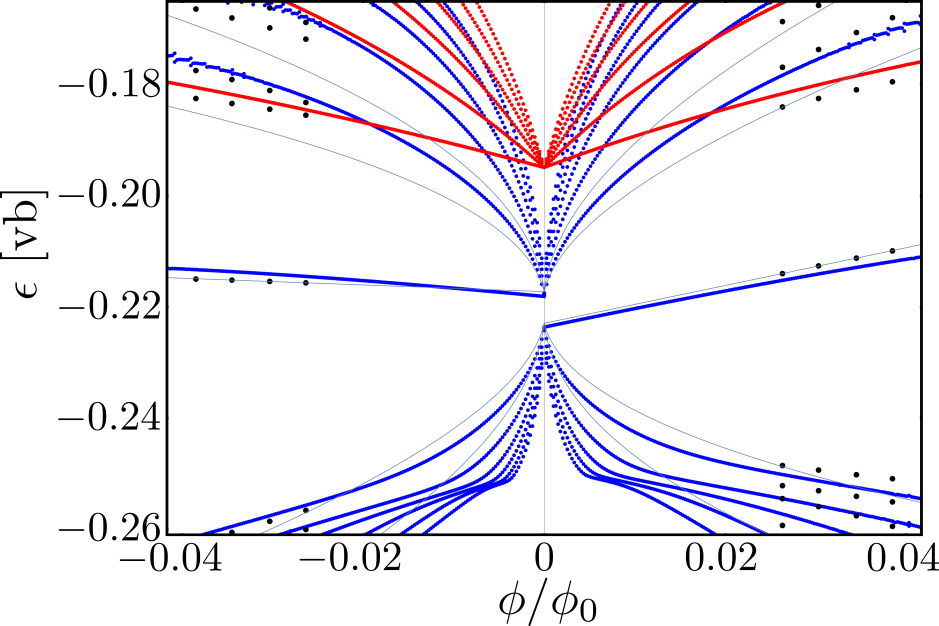}
\caption{
The Landau level spectra of the valence band secondary Dirac point (sDP) the perturbation $u_{0,1,3}=\{0.032, -0.063, -0.055\}$,
obtained from either Hamiltonian~\eqref{eq:sDPH} for the $\kappa'$ and $\kappa$ corners (red and blue dots) or Hamiltonian \eqref{eqn:2x2Hamiltonian} (black dots).
The grey solid lines are the Landau levels of the effective Hamiltonian, Eq.~\eqref{eqn:2x2SecondaryDPHamiltonian}, using fitted parameters $E_0=-0.22vb$, $\Delta_{\kappa}=-0.0057vb$, $\tilde{v}=0.38v$,
$\bar{M}=-0.97 e v / b $, and $\tilde{M}=-1.45e v /  b$.
}
\label{fig:sDPLL}
\end {figure}
Here we give an analytical description of the almost gapless sDP found at the corner superlattice Brillouin zone in the valence band in the top row of Fig.~\ref{fig:magnetic_minibands}.
To do this, we note that zone folding using Bragg vectors $\vect b_m$ brings together three degenerate plane wave states $|\zeta\!\vect \kappa\rangle$,  $|\zeta\!(\vect \kappa +\vect b_2)\rangle$ and  $|\zeta\!(\vect \kappa+\vect b_1)\rangle$to each of the two inequivalent corners of the moir\'e Brillouin zone where $\zeta=\pm1$, $\vect \kappa=(\vect b_4+\vect b_5)/3$,  and,
\begin{align}
|\zeta \vect p \rangle
=\frac{1}{\sqrt{2}}\left( \begin{array}{c} 1 \\ -\alpha e^{2 i \varphi }\end{array} \right)e^{i \zeta \vect p.\vect{r}},
\label{eq:EDiracBLG}
\end{align}
with $\varphi$ the polar angle of momentum $\zeta \vect p$, and $\alpha=\pm1$ the band index.
Using $\vect k\cdot \vect p$ theory, the vicinity of each moir\'e Brillouin zone corner can then be described using an effective Hamiltonian acting on a three component vector of smoothly varying envelope functions, written in a basis of the standing wave states,
\begin{align}
\label{eq:standing_waves}
&\Phi_1\!=\!\frac{1}{\sqrt{3}}\left(  |\zeta\!\vect \kappa\rangle+|\zeta\!(\vect \kappa +\vect b_2)\rangle+|\zeta\!(\vect \kappa+\vect b_1)\rangle \right),\\
&\Phi_2\!=\!\frac{1}{ \sqrt{3}} \left(e^{i\pi } |\zeta\!\vect \kappa\rangle+ e^{\text{-}i\frac{\pi}{3} }|\zeta\!(\vect \kappa +\vect b_2)\rangle+ e^{i\frac{\pi}{3} }|\zeta\!(\vect \kappa+\vect b_1)\rangle
\right),\nonumber\\
&\Phi_3\!=\!\frac{1}{\sqrt{3}}\left(e^{\text{-}i\pi }|\zeta\!\vect \kappa\rangle +  e^{ i\frac{\pi}{3} } |\zeta\!(\vect \kappa +\vect b_2)\rangle +  e^{\text{-}i\frac{\pi}{3} }|\zeta\!(\vect \kappa+\vect b_1)\rangle  \right). \nonumber
\end{align}
Using basis \eqref{eq:standing_waves} the $\vect k \cdot \vect p$ Hamiltonian for the $K$-valley is diagonal exactly at the Brillouin zone corner,
\begin{align} \label{eq:sDPH}
&\op{H}_K(\zeta, \vect{p}, \vect{B})    \\
&=\!\!\!\left(\!\begin{array}{ccc}\op{\epsilon}_0\!+\!\epsilon_1\!\! & \tilde{v} \op{\pi}^{\dagger }
&  \tilde{v} \op{\pi}
  \\ \tilde{v} \op{\pi} & \!\!\op{\epsilon}_0\!+\!\epsilon_2 \!\!& \!-\!\tilde{v} \op{\pi}^{\dagger } \\
   \tilde{v} \op{\pi}^{\dagger } &  \!-\!\tilde{v} \op{\pi} & \!\!\op{\epsilon}_0\!+\!\epsilon_3
  \end{array}\!\right)\!
\nonumber \\
&\,\,+\tilde{v}\!\left(\!
\begin{array}{ccc}
0 & \!\!\!\!-\!(\eta_1 \!+\!\eta_2 \!+\!\eta_3 )  \op{\pi}^\dagger\!\! & \!\!(\eta_2 \!-\!\eta_3 )  \op{\pi} \\
 \!-\!(\eta_1 \!+\!\eta_2 \!+\!\eta_3 )  \op{\pi} \!\!& 0 & \!\!\!-\!(\eta_1 \!+\!2 \eta_3 )  \op{\pi}^\dagger \\
 (\eta_2 \!-\!\eta_3 ) \op{\pi}^\dagger\!\! & \!\!\!-\!(\eta_1 \!+\!2 \eta_3 ) \op{\pi}\!\! & 0 \\
\end{array}
\!\right)\!;\nonumber\\
&\epsilon_1 =2\Re W, \
\epsilon_2 =- \Re W  + \sqrt{3} \Im W, \
\epsilon_3 =- \Re W  - \sqrt{3} \Im W, \nonumber \\
&W\!\approx\!\frac{ vb}{2}(u_0\!+\!i \zeta u_3)
\!+\!\frac{v^3 b^3}{6\gamma_1^2} e^{\frac{2i \pi }{3}}(u_0\!-\!i \zeta u_3)
\!+\!  \frac{\alpha\zeta v^2 b^2}{   \sqrt{3}\gamma_1} e^{\frac{\text{-}2i \pi}{3 }} u_1,
\nonumber
\\
& \op{\epsilon}_0\!=\!\frac{\alpha v^2 b^2}{3 \gamma _1} \!+\!\frac{\alpha v^2}{\gamma _1}\op{p}^2,
\quad
\tilde{v} \!=\!\zeta \alpha \frac{ b v^2}{\sqrt{3}\gamma_1},
\quad
\eta_3=\frac{1}{2} \sqrt{3} \zeta  \mu_1
\nonumber \\
&\eta_1=\frac{vb \left(\sqrt{3} \zeta  \mu_3 +\mu_0 \right)}{2 \gamma_1},\eta_2=\frac{vb \left(\mu_0 -\sqrt{3} \zeta  \mu_3 \right)}{2 \gamma_1},\nonumber
\end{align}
where $\Re$ and $\Im$ denote the real and imaginary parts.
The corresponding Hamiltonian for the $K'$-valley is obtained using time reversal symmetry, $\op{H}_K( \zeta, \vect{p}, \vect{B})$=$\op{H}^*_{K'}( \text{-}\zeta, \text{-}\vect{p}, \text{-}\vect{B})$.

Figure \ref{fig:sDPLL} shows the LL spectra of Hamiltonian \eqref{eq:sDPH} using blue(red) dots for the $-\kappa$($\kappa$) corner, where the signature of an almost gapless sDP is evident at $\epsilon \approx -0.22\,vb$ for the $-\kappa$ corner.
To calculate this spectra, we diagonalized the Heisenberg matrix of Hamiltonian \eqref{eq:sDPH} in a basis consisting of the products of magnetic oscillator functions \eqref{eq:mag_oscil_funcs} with the standing waves states \eqref{eq:standing_waves}, $\varphi_n(k_2) \Phi_i$,  where we take $i=1,2,3$, and $n=0\cdots n_{\text{max}}$, where is $n_{\text{max}}$ sufficiently large that the resulting spectrum converges. Any eigenstates with a large weight on high-index magnetic oscillator functions, $\varphi_n(k_2)$ with  $n\gtrsim (b  \lambda_B )^2  $, are discarded because they violate the $\vect k\cdot \vect p$ approximation used to construct Hamiltonian \eqref{eq:sDPH}. Also, to improve the comparison of its LL spectra with that of Hamiltonian \eqref{eqn:2x2Hamiltonian}, we calculate the moir\'e perturbation correction to the band-edge energies, $\epsilon_{1,2,3}$, using a higher order of perturbation theory than is explicit in Eq.~\eqref{eq:sDPH}.

The result of numerical diagonalization of the two band Hamiltonian \eqref{eqn:2x2Hamiltonian} is displayed using black dots in Fig.~\ref{fig:sDPLL}.
The two spectra agree well, confirming that Eq.~\eqref{eq:sDPH} is a good description of the sDP.
Moreover, we can diagonalize Hamiltonian \eqref{eq:sDPH} for arbitrarily small magnetic fields, where the size of the Heisenberg matrix of magnetic Bloch functions \eqref{eq:bloch_wf} needed to diagonalize Hamiltonian \eqref{eqn:2x2Hamiltonian} becomes prohibitively large.

The energy of the``zeroth'' LL originating from the sDP in Fig.~\ref{fig:sDPLL} depends on the magnetic field, due to a non-zero magnetic momentum generated from the influence of a third band (residing mainly on $\Phi_3$ for Hamiltonian \eqref{eq:sDPH}) that mixes with the LLs of the first two bands.
For the parameter set $u_{0,1,3}=\{0.032, -0.063, -0.055\}$, this third band is separated by a large gap from the sDP, which permits the approximate two band description for the sDP,
\begin{align} \label{eqn:2x2SecondaryDPHamiltonian}
&\op{H}_{\text{sDP}}^{\text{eff}}\!=\! E_0
\!+\!\frac{\Delta_\kappa}{2}\sigma_3
\!+\! \tilde{v} \op{p}\cdot \vect{\sigma}
\!-\! \left(\bar M +\tilde M\sigma_3 \right) \!B.
 \end{align}
where the parameters for the energy shift $E_0$, gap $\Delta_\kappa$,
effective velocity $\tilde{v}$, and magnetic momentums $M$ and $\tilde M$ can be fitted to the result of numerically diagonalizing Eq.~\eqref{eqn:2x2Hamiltonian}.
Such a fitting is illustrated in Fig.~\ref{fig:sDPLL} using the grey lines, and produces the best accuracy for low indexed LLs.

\section{Magnetic miniband spectrum and compressibility of 2D electron liquid in BLG-hBN heterostructures}\label{S:map}
\begin{figure}[htbp]
\centering
\includegraphics[width=.48 \textwidth]{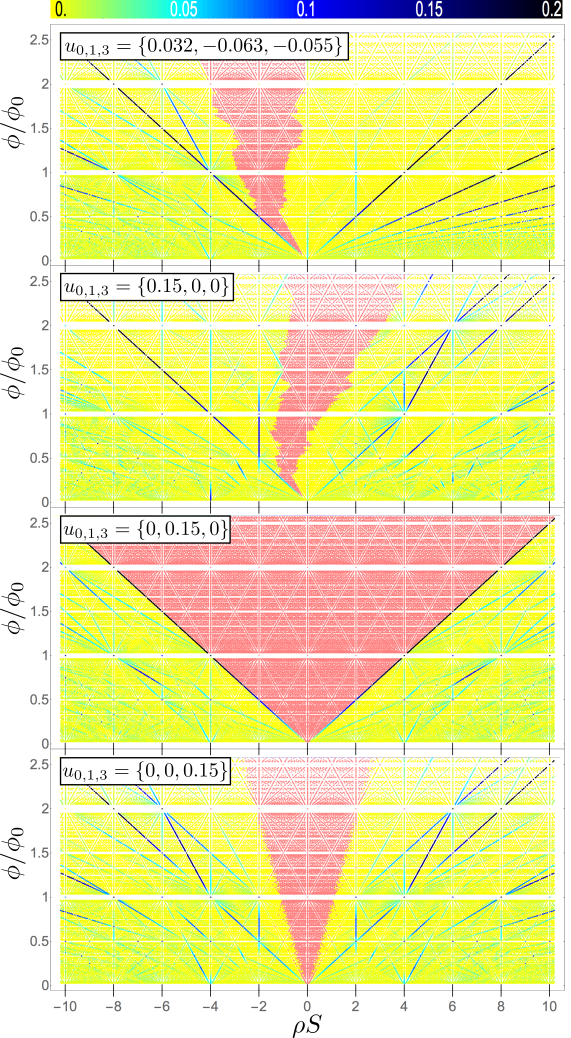}
\caption{
Density-magnetic field diagram showing incompressible
electron states in the magnetic minibands,
calculated for the same parameters as Fig.~\ref{fig:magnetic_minibands}.
Larger gaps are shown using a darker color, yellow represents a vanishing gap, and red represents the particular gapless states associated with the zero-energy Landau level.
}
\label{fig:filling_factors}
\end {figure}
Fig.~\ref{fig:filling_factors} shows the maps of incompressible states for each parameter set in Fig.~\ref{fig:magnetic_minibands}, created by filling the corresponding magnetic miniband spectrum with electrons.
A larger gap is depicted with a darker color, resulting in lines, the gradient of which corresponds to the filling factor.
For the regions shown in yellow, the gap is negligibly small as a result of incomplete filling
of a magnetic miniband, each of which can accommodate a density $\delta \rho = 1 / (\pi p \lambda_B^2)$ of electrons (including spin degeneracy). In contrast, the unperturbed zero-energy Landau level present in the $K^{\prime}$ valley accommodates electron density of $\delta \rho_0 = 1 / (\pi \lambda_B^2)$, creating a large gapless region portrayed in red. Because of its large electron capacity at exactly zero energy, the presence of this zero-energy mode should be clearly visible in capacitance measurements.
Electron-hole symmetry is displayed in the plots for
$u_{0,1,3}=\{0,0.15,0\}$ (lower left),
and $u_{0,1,3}=\{0,0,0.15\}$ (lower right), whereas it is absent in the upper two plots.

Many of the largest gaps in the magnetic spectrum plots (Fig.~\ref{fig:magnetic_minibands}) occur between the zeroth and first LLs of Dirac points (whether original or secondary).  In the map of incompressible states, these large gaps are manifest as a solid blue line, which intersect the $B=0$ axis at a density $\rho S = 4m$, where $m=0$ for the main Dirac point or $m=\pm1$ for an sDP at the edge of the first conduction/valence miniband.
For example, there is a (gapped) sDP in the valence band of the magnetic spectrum for the moir\'e perturbation $u_{0,1,3}=\{0.032, -0.063, -0.055\}$ (upper panel in Fig.~\ref{fig:magnetic_minibands}).
In the upper panel of Fig.~\ref{fig:filling_factors}, the gap between the two zeroth LLs of this sDP is represented as a blue vertical line, which starts from ($\rho S=-4$, $B=0$), whereas the two gaps between the zeroth and first LLs are shown as two tilted blue lines, with gradients of $\pm2$.

\section{Conclusions}

In summary, we have shown that the presence of the {\hbn} substrate lifts the valley degeneracy of bilayer graphene, producing different magnetic Hofstadter's butterflies in each valley.
The zero-energy Landau level located on the layer furthest from the {\hbn} substrate remains unaffected by the moir\'e perturbation, which makes the BLG/{\hbn} spectrum unique in comparison to other known magnetic spectra, for which all Landau levels split into sub-bands.
We investigated the influence of different possible characteristics of moir\'e perturbation including an electrostatic potential which
does not distinguish between the two carbon sublattices, a sublattice-asymmetric part of the potential, and a spatial modulation of the nearest-neighbor carbon-carbon hopping amplitude, and we identified how they give rise to different features in the electronic spectra including
gapped or overlapping bands, or bands connected by a secondary Dirac point.
In addition to determining the fractal electronic spectra by numerical diagonalization of a model Hamiltonian, we also derived simple effective Hamiltonians to describe low-energy features in the magnetic minibands and to describe the secondary Dirac point.
Finally, we showed how gaps in the fractal energy spectrum lead to the formation of incompressible states that may be observed under a variation of carrier density and magnetic field.

\begin{acknowledgments}
We thank A.~K.~Geim, I.~Aleiner and M.~Koshino for useful discussions. This work was funded by the EU Flagship Project, EPSRC Science and Innovation Award, EPSRC Grant EP/L013010/1, and ERC Synergy Grant Hetero2D.
\end{acknowledgments}

\end{document}